\documentclass{moriond}
\usepackage{amssymb}
\usepackage{mathtools}
\newcommand{\Photo}{}

\begin{document}
\vspace*{4cm}
\title{Probes of Heavy Sterile Neutrinos}

\author{Patrick D. Bolton$^1$, Frank F. Deppisch$^{2}$\footnote{Corresponding Author: f.deppisch@ucl.ac.uk}, P. S. Bhupal Dev$^3$}
\address{$^{1)}$SISSA, International School for Advanced Studies and INFN, Sezione di Trieste,\\Via Bonomea 265, I-34136 Trieste, Italy\\
$^{2)}$Department of Physics and Astronomy, University College London,\\Gower Street, London WC1E 6BT, U.K.\\
$^{3)}$Department of Physics and McDonnell Center for the Space Sciences,\\Washington University, St. Louis, MO 63130, U.S.A.
}

\maketitle\abstract{We review probes of heavy sterile neutrinos, focusing on direct experimental searches and neutrinoless double beta decay. Working in a phenomenological parametrization, we emphasize the importance of the nature of sterile neutrinos in interpreting neutrinoless double beta decay searches. While current constraints on the active-sterile neutrino mixing are already stringent, we highlight planned future efforts that will probe regimes motivated by the lightness of active neutrinos.}

\section{Introduction}
The lack of right-handed neutrino states $N_j$ is an intriguing feature of the Standard Model (SM). Their presence is not strictly required, unlike for the other SM fermions, as we only observe neutrinos through their left-handed SM interactions, and the small but finite masses could be understood with the active left-handed states $\nu_\ell$ only, assuming their Majorana nature. This does not mean that right-handed neutrinos do not exist, though: Because they would be \emph{sterile}, i.e., uncharged under the SM gauge interactions, they participate only in the Yukawa interaction $-y_\nu^{\ell j} \bar L_\ell\cdot H N_j$ with a left-handed lepton doublet and the Higgs doublet. If small, $y_\nu \sim m_\nu / v \lesssim 10^{-12}$ ($v = 174$~GeV), this would generate light \emph{Dirac} neutrino masses $m_\nu \lesssim 0.1$~eV seen in oscillations and constrained by absolute mass searches such as tritium decay and cosmological observations.

The introduction of sterile, right-handed neutrino states opens up the possibility of \emph{lepton number violation}. Whereas total lepton number is an accidental symmetry in the SM, sterile neutrinos can have a \emph{Majorana} mass term, $-\frac{1}{2}M^{ij} \bar N_i^C N_j$, violating lepton number by two units, unless it is explicitly forbidden by an additional symmetry beyond the SM gauge ones. Such a Majorana mass term is fairly unconstrained from a theoretical point of view, as it is not connected to the SM electroweak symmetry breaking. It can in principle have any scale from eV and below to scales far above the SM. 

With both the Yukawa and a heavy sterile Majorana mass term present, the neutrino spectrum consists of three (mostly) active light neutrinos and two or more (mostly) sterile neutrinos, all having Majorana character. This is the celebrated \emph{seesaw mechanism} (of type 1), connecting the light neutrino masses to the heavy Majorana mass scale $m_N$ as $|m_\nu| \sim (y_\nu v)^2/m_N = |V_{\ell N}|^2m_N$, applicable if $y_\nu v \ll m_N$. Here, $|V_{\ell N}| = y_\nu v/m_N$ is the active-sterile mixing strength. Its main phenomenological consequence is that it generates suppressed charged and neutral current interactions between the sterile state $N$ and a SM charged lepton or neutrino of flavour $\ell = e,\mu,\tau$. The lightness of active neutrinos can thus be explained by either making the sterile neutrinos heavy or the Yukawa coupling weak. The admixture of a sterile neutrino with an active one, $N^\text{mass} \sim N - V_{\ell N} \nu_\ell$, is in either case small,
\begin{align}
    |V_{\ell N}| = \sqrt{\frac{m_\nu}{m_N}} 
                   \lesssim 10^{-6} \sqrt{\frac{100~\text{GeV}}{m_N}}.
\end{align}

There is a third way of keeping the active neutrinos light: If lepton number were to be conserved in the sterile neutrino sector, no light masses are generated. This is not only achieved in the limits $m_N\to 0$ and $m_N\to\infty$ but also if pairs of sterile states form Dirac particles themselves. By violating lepton number symmetry slightly, e.g., through a Majorana mass $\mu \ll m_N$, light neutrino masses are generated, $|m_\nu| \sim (y_\nu v/m_N)^2 \mu = |V_{\ell N}|^2 \mu$, while the heavy sterile neutrinos form quasi-Dirac pairs with a small mass splitting $\Delta m_N \sim \mu$. This concept applies to extended scenarios such as the \emph{inverse seesaw mechanism} where the scale of lepton number breaking $\mu$ is decoupled from the heavy neutrino mass $m_N$.

This proceedings report is based on our paper comparing the sensitivity of direct searches with that of neutrinoless double beta ($0\nu\beta\beta$) decay~\cite{Bolton:2019pcu}. It utilizes a phenomenological pa\-ra\-metrization describing the general mixing of two heavy sterile neutrino states with one generation of active neutrino where the purely Majorana and Dirac scenarios can be understood as limiting cases.

\section{Current Constraints and Future Sensitivities}
\begin{figure}[t!]
	\centering
	\includegraphics[width=0.79\textwidth]{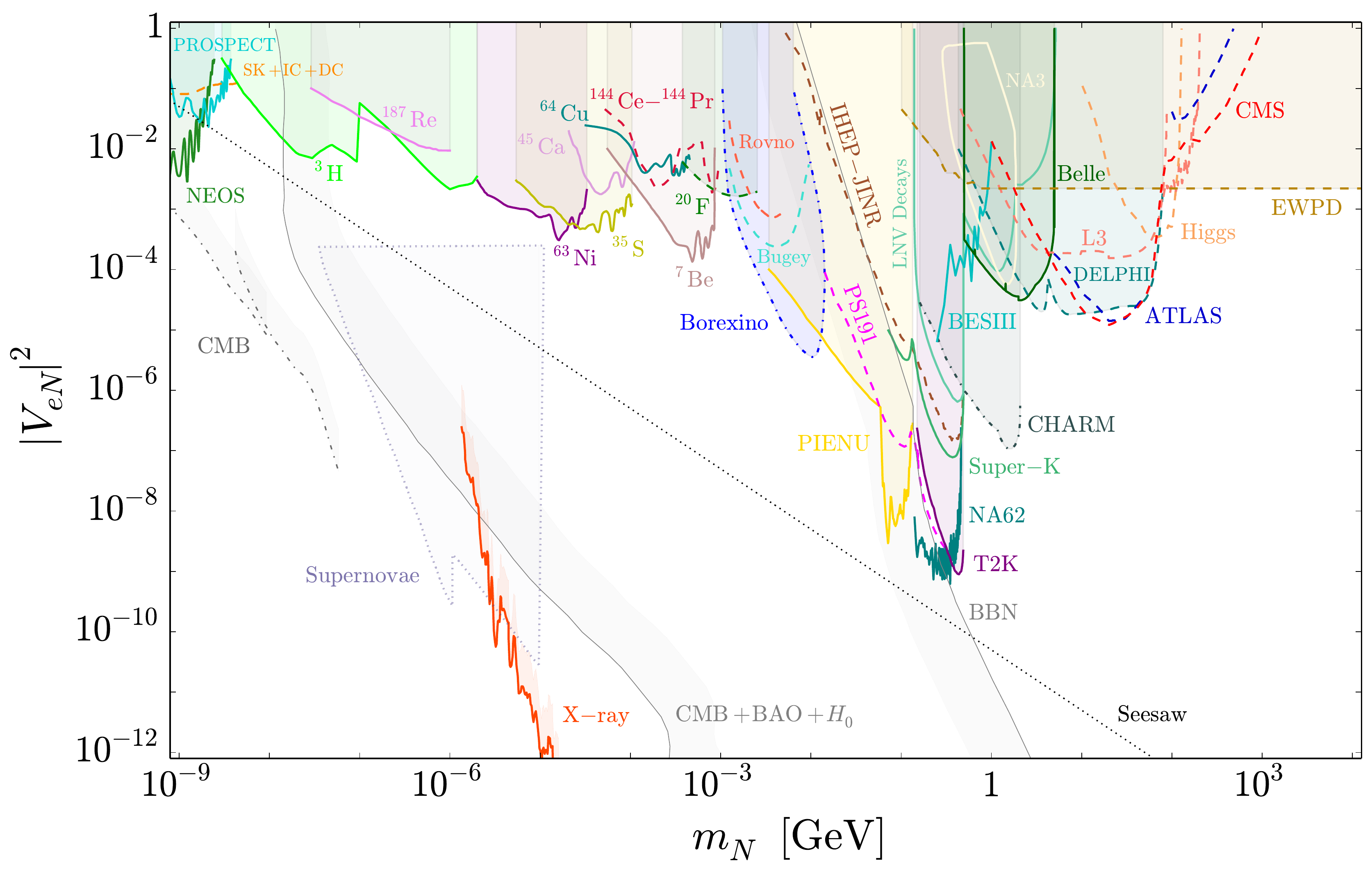}
	\caption{Current constraints on the active-sterile mixing strength $|V_{eN}|$ as a function of the sterile neutrino mass $m_N$. Adapted from the companion paper, with detailed descriptions of the various probes therein and in the accompanying website \url{www.sterile-neutrino.org}.}
	\label{fig:current}
\end{figure}

We briefly describe the different classes of probes. For details we refer the reader to our paper~\cite{Bolton:2019pcu}, its accompanying website \url{www.sterile-neutrino.org} and other recent literature~\cite{Abdullahi:2022jlv}. We concentrate on the mixing strength $|V_{eN}|$ of a heavy sterile neutrino with electron flavour as this is the one relevant for $0\nu\beta\beta$ decay. Current constraints on $|V_{eN}|$ as a function of the heavy sterile neutrino mass $m_N$ are shown in Fig.~\ref{fig:current}, over the broad mass range $0.1~\text{eV} < m_N < 10$~TeV. The diagonal line labelled \textbf{Seesaw} indicates the mixing strength expected in canonical seesaw, $m_\nu = |V_{eN}|^2 m_N$, generating a light neutrino mass $m_\nu = 0.05$~eV. Reaching it may be considered the ultimate goal for sterile neutrino searches, though, both larger (e.g., in inverse seesaw models) and smaller (where other contributions dominate the generation of light masses) mixing strengths are possible.

\begin{figure}[t!]
	\centering
	\includegraphics[width=0.79\textwidth]{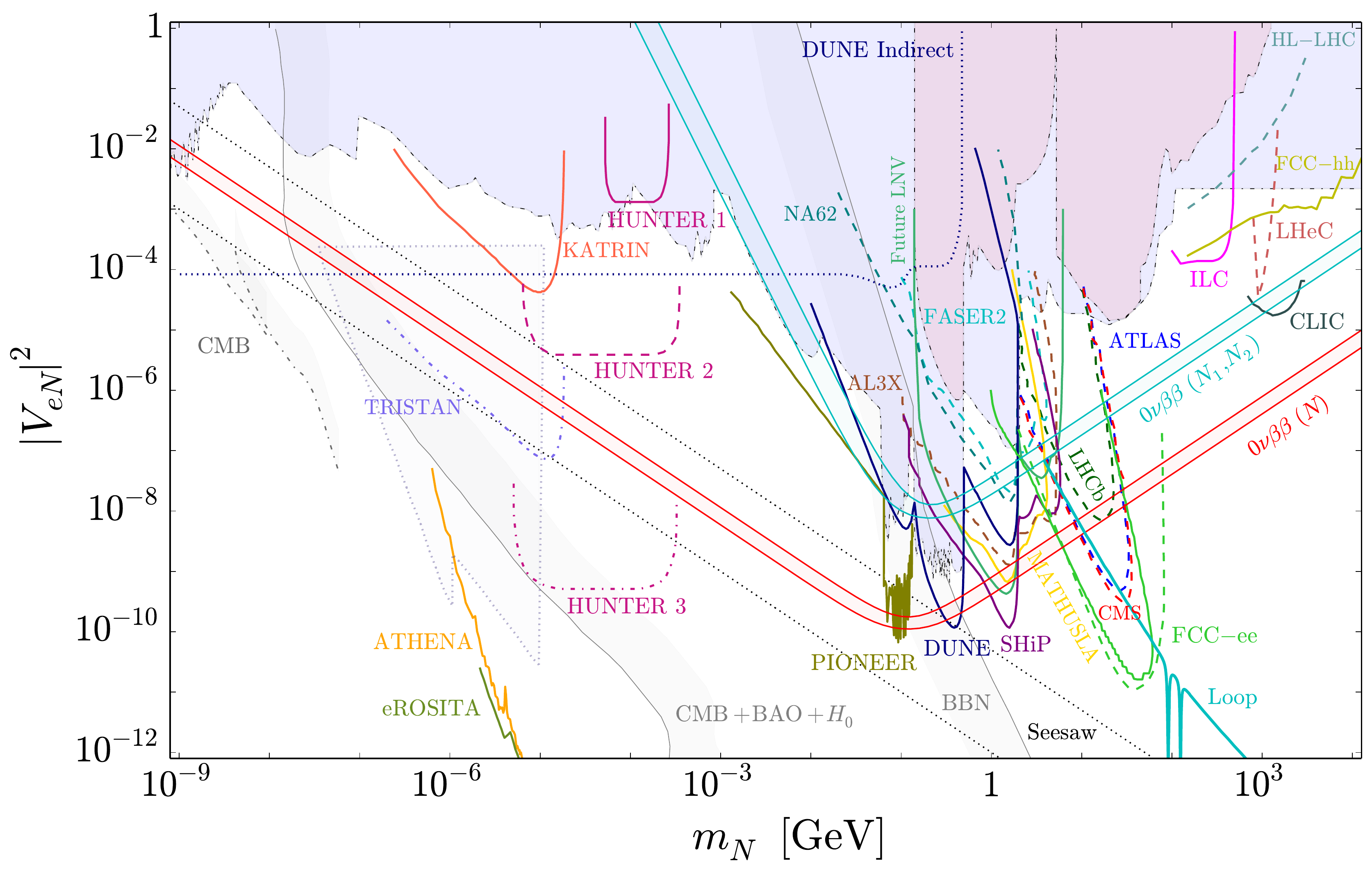}
	\caption{As Fig.~\ref{fig:current}, but showing the projected sensitivities of future searches (open curves), including $0\nu\beta\beta$ decay, over the existing constraints (shaded regions). The region in light blue is disallowed for both Dirac and Majorana sterile neutrinos whereas the region in light red applies only for Majorana neutrinos. The $0\nu\beta\beta$ decay sensitivities are for a heavy sterile Majorana (red) and quasi-Dirac neutrino (teal) for the half-life $T^{0\nu}_{1/2} = 10^{28}$~yr in $^{76}$Ge. Adapted from the companion paper, with detailed descriptions of the various probes therein and in the accompanying website \url{www.sterile-neutrino.org}.}
	\label{fig:future}
\end{figure}

It may be surprising that sterile neutrinos, having no inherent gauge charges, are constrained from so many directions, but the active-sterile mixing causes the sterile neutrinos to participate in charged-current and neutral-current SM interactions, albeit suppressed by~$|V_{\ell N}|$. As opposed to the light active neutrinos, heavy sterile neutrinos can decay, either promptly or with a macroscopic proper decay length,
\begin{align}
    L_N \approx 25~\text{mm}
    \cdot\frac{10^{-10}}{|V_{\ell N}|^2}
    \cdot\left(\frac{10~\text{GeV}}{m_N}\right)^2,
\end{align}
where this approximation is roughly valid for $1~\text{GeV} \lesssim m_N \lesssim m_W$. This improves detectability and especially the \emph{long-lived particle (LLP)} signature, in combination with high intensity production mechanisms, allows probing sterile neutrinos with very small mixing strengths, close to the canonical seesaw floor. If a heavy sterile neutrino were to be discovered near this line, it would most likely be of Majorana nature. Most probes, such as direct searches, where sterile neutrinos are produced on-shell, are sensitive to both Majorana and Dirac neutrinos. Only those probes relying on a total lepton number violating signal require a Majorana sterile neutrino, c.f. Fig.~\ref{fig:future}, where corresponding limits are highlighted in red.

The strong reach of high-luminosity, displaced-vertex searches is especially apparent in Fig.~\ref{fig:future}, which shows sensitivities of proposed experiments (open curves) in comparison with current constraints (shaded regions).

\subsection{High-energy Colliders}
Sterile neutrinos are produced in high-energy collisions through charged and neutral currents such as $pp\to W^+\to e^+ N$ and $pp\to Z\to \nu N$. For example, the proposed \textbf{FCC-ee} will be a powerful $Z$ factory with low background for displaced vertices in a large detector volume. 

\subsection{Meson Decays and Beam-Dump Experiments}
Likewise, beam-dump experiments and meson factories produce a large number of mesons that subsequently decay to heavy sterile neutrinos, e.g., $K^+ \to e^+ N$. The most sensitive direct limits are currently from \textbf{NA62} and the long baseline neutrino oscillation experiment \textbf{T2K}, exploiting this route in existing facilities. Future searches such as \textbf{SHiP} can be purpose-built with optimized LLP detectors. 

\subsection{Beta Decays and Nuclear Processes}
Sterile neutrinos mixing with electron-flavour and masses $m_N \lesssim 1$~MeV can be produced in nuclear beta decays and other weak nuclear processes. They will suppress the rate and can produce a detectable kink in the decay spectrum. Strong advancements in such searches are expected, with the recent BeEST experiment~\cite{Friedrich:2020nze} improving previous limits by almost two orders of magnitude ({\boldmath$^7$}\textbf{Be}). Future searches by the same collaboration and, e.g., the \textbf{HUNTER} proposal~\cite{Martoff:2021vxp} aim to go below the region disfavoured by cosmological considerations, probing a parameter space where the sterile neutrino can be a viable Dark Matter candidate.

\subsection{Active-Sterile Neutrino Oscillations}
Sterile neutrinos can be produced through oscillations with the active states. Persistent anomalies around the squared mass difference $\Delta m_{14}^2 = m_N^2 - m_\nu^2 \approx 1~\text{eV}^2$ hint at the presence of a sterile neutrino around this scale. Interpreted conservatively, oscillation experiments provide constraints on light eV-scale sterile neutrinos. Heavier sterile neutrinos can also be probed through the resulting deficit of active neutrinos detected, though this requires a very good understanding of the absolute neutrino flux. For DUNE, this is indicated by the contour labelled \textbf{DUNE Indirect}~\cite{Carbajal:2022zlp}. 

\subsection{Electroweak Precision Data and Indirect Laboratory Constraints}
Likewise, any mixing with sterile neutrinos means that the active neutrino mixing matrix itself is non-unitary. This is visible in charged current and neutral current processes, altering electroweak precision data (\textbf{EWPD}) observables.

\subsection{Cosmological and Astrophysical Constraints}
Sterile neutrinos are produced in the early universe through scattering or oscillation. If decaying at times later than $\sim 1$~s, they affect the abundance of primordial elements in big bang nucleosynthesis (\textbf{BBN}). Decays of longer-lived sterile neutrinos will inject radiation degrees of freedom, i.e., light active neutrinos, which are strongly constrained. Furthermore, the loop-induced decay $N\to\nu\gamma$ is detectable in astrophysical \textbf{X-ray} observations. Quasi-stable sterile neutrinos will act as Dark Matter and must not overclose the universe (\textbf{CMB+BAO+}{\boldmath$H_0$}). Taken at face value, such considerations disfavour much of the parameter space $m_N \lesssim 1$~GeV accessible in laboratory experiments. They can be relaxed in extended scenarios, e.g., where sterile neutrinos decay to a dark sector modifying their lifetime.

\subsection{Neutrinoless Double Beta Decay}
\label{sec:0vbb}

The most sensitive probe of the Majorana nature of light active neutrinos is $0\nu\beta\beta$ decay~\cite{Agostini:2022zub}. Observing this rare nuclear process in select isotopes such as $^{76}\text{Ge} \to {}^{76}\text{Se} + e^- e^-$, is only possible if total lepton number is violated. It would prove that the light active neutrinos are Majorana fermions. In addition, $0\nu\beta\beta$ decay is sensitive to other exotic sources of lepton number violation, typically at or below the $\mathcal{O}(10)$~TeV scale~\cite{Doi:1981,Cirigliano:2017djv,Graf:2018ozy,Deppisch:2020ztt}. 

This includes the exchange of sterile neutrinos. In particular, $0\nu\beta\beta$ decay is highly sensitive to heavy Majorana neutrinos. In this case, the decay half-life $T_{1/2}^{0\nu}$ for $m_N \gtrsim 100$~MeV is approximately given by
\begin{align}
\label{eq:0vbb:half-life-heavy}
	\frac{10^{28}~\text{yr}}{T_{1/2}^{0\nu}} \approx \left(\frac{|V_{eN}|^2}{10^{-9}}\cdot\frac{1~\text{GeV}}{m_N}\right)^2.
\end{align}
For lighter sterile neutrinos, $m_N \lesssim 100$~MeV, the rate is proportional to $m_N^2$,
\begin{align}
\label{eq:0vbb:half-life-light}
	\frac{10^{28}~\text{yr}}{T_{1/2}^{0\nu}} \approx \left(\frac{|V_{eN}|^2}{10^{-9}}\cdot\frac{m_N}{15~\text{MeV}}\right)^2,
\end{align}
The above approximations use nuclear matrix elements for the isotope $^{76}$Ge~\cite{Deppisch:2020ztt}. The behaviour changes around the nuclear momentum scale $\approx 100$~MeV of $0\nu\beta\beta$ decay, and at the crossover the momentum dependence should be accounted for more carefully~\cite{Babic:2018ikc,Dekens:2020ttz}. 

The above sensitivities are normalized with respect to the half-life $T^{0\nu}_{1/2}(^{76}\text{Ge}) = 10^{28}$~yr. This is the projected sensitivity of LEGEND-1000~\cite{Zsigmond:2020bfx}, one among a range of proposed experiments~\cite{Agostini:2022zub} mainly aiming to probe the light active Majorana neutrino parameter space for an inverted mass ordering. Current experimental limits are of the order $T^{0\nu}_{1/2} \gtrsim 10^{26}$~yr~\cite{PhysRevLett.117.082503,GERDA:2020xhi}.

In Fig.~\ref{fig:future}, the future sensitivity to Majorana sterile neutrinos is given by the red band labelled {\boldmath$0\nu\beta\beta~(N)$}, where the width indicates the theoretical uncertainty from nuclear matrix elements. This includes the potential quenching of the axial nuclear coupling strength~\cite{Deppisch:2016rox}. 

Future $0\nu\beta\beta$ decay experiments can reach sensitivities $|V_{eN}|^2 \approx 2\times 10^{-10}$ to $10^{-9}$, in the regime $10~\text{MeV}\lesssim m_N \lesssim 1$~GeV. This is in the range expected for the canonical seesaw with $m_N \approx 100$~MeV, and also comparable to future direct searches in this mass window such as \textbf{PIONEER}~\cite{PIONEER:2022yag} and \textbf{DUNE}. 

For masses and mixing outside this range, the nominal sensitivity to heavy sterile Majorana neutrinos is still strong, but as mentioned above, the light masses naturally require that sterile neutrinos are quasi-Dirac states with an associated small mass splitting that suppresses $0\nu\beta\beta$ decay. For example, with a pair of quasi-Dirac sterile neutrino states of average mass $m_N$ and relative mass splitting $\delta_N = \Delta m_N/m_N$, Eq.~\eqref{eq:0vbb:half-life-heavy} is modified to 
\begin{align}
	\label{eq:0vbb:half-life-heavy-quasi-dirac}
	\frac{10^{28}~\text{yr}}{T_{1/2}^{0\nu}} 
	\approx \left(\frac{\delta_N}{10^{-2}}
	\cdot\frac{|V_{eN}|^2}{10^{-7}}
	\cdot\frac{1~\text{GeV}}{m_N}\right)^2.
\end{align}
Larger masses $m_N \gtrsim 10$~GeV and splittings are in principle possible but they require a fine-tuned cancellation of the induced loop contribution to the light neutrino masses~\cite{Mitra:2011qr,Lopez-Pavon:2012yda,Bolton:2019pcu}. Direct searches looking for lepton number conserving signals do not have such a suppression and they can probe purely Dirac-type sterile neutrinos.

The above discussion assumes that the sterile neutrino contribution saturates the $0\nu\beta\beta$ decay sensitivity, but other mechanisms may be present. Most directly, the light active neutrinos (if Majorana) will induce the effective $0\nu\beta\beta$ mass $m_{\beta\beta}$ that destructively interferes with sterile neutrinos for $m_N \lesssim 100$~MeV if these participate in the seesaw mechanism. For $m_N \ll 100$~MeV, light and heavy neutrino contributions will cancel to zero in this case.

The effect of all three types of suppression is indicated in Fig.~\ref{fig:future} by the teal band labelled {\boldmath$0\nu\beta\beta~(N_1, N_2)$}: A quasi-Dirac sterile neutrino nature with $\delta = 10^{-2}$ leads to an overall reduction of sensitivity with respect to the red Majorana band, the interference with the light active neutrino contribution (with $m_\nu = m_{\beta\beta} = 10^{-3}$~eV, also indicated by the lower diagonal \textbf{Seesaw} line) induces a steeper slope for $m_N < 100$~MeV, and strong loop corrections to the light neutrino masses of 10\% or more are present to the top and right of the line labelled~\textbf{Loop}.

\section{Discussion}
Given their curious absence from the SM and their importance as the potential origin of the light neutrino masses, it is only right that sterile neutrinos are being probed in a large number of experiments and observations. We have briefly highlighted the main approaches to search for sterile neutrinos in the experimentally accessible range $1~\text{eV} \lesssim m_N \lesssim 10$~TeV. Beyond this regime, we must primarily resort to theoretical considerations, e.g., the stability of the Higgs potential modified by the Yukawa couplings of the sterile neutrinos.

Apart from their connection to neutrino masses, sterile neutrinos may play a crucial role in explaining the matter-antimatter asymmetry of the universe through their participation in various \emph{leptogenesis} scenarios. This provides a major, additional motivation to search for sterile neutrinos, especially in the range $1~\text{GeV} \lesssim m_N \lesssim 100$~GeV.

Lastly, sterile neutrinos may have additional interactions beyond the ones induced by the active-sterile mixing, leading to other portals such as transition magnetic moments~\cite{Bolton:2021pey} or exotic gauge interactions~\cite{Liu:2022kid,Padhan:2022fak}.

\section*{Acknowledgments}
F.F.D. acknowledges support from the UK Science and Technology Facilities Council (STFC) via the Consolidated Grants ST/P00072X/1 and ST/T000880/1. P.D.B. has received support from the European Union’s Horizon 2020 research and innovation programme under the Marie Skłodowska-Curie grant agreement No. 860881-HIDDeN. The work of B.D. is supported in part by the US Department of Energy under Grant No. DE-SC0017987.

\section*{References}
\bibliography{bibliography}
\end{document}